\documentclass[conference]{IEEEtran}
\IEEEoverridecommandlockouts
\usepackage{cite}
\usepackage{amsmath,amssymb,amsfonts}
\usepackage{algorithmic}
\usepackage{graphicx}
\usepackage{textcomp}
\usepackage{xcolor}
\usepackage{hyperref}
\usepackage[capitalise]{cleveref}
\usepackage[acronym]{glossaries}
\usepackage{xspace}
\usepackage{url}
\usepackage{subcaption}
\usepackage[colorinlistoftodos,prependcaption,textsize=tiny]{todonotes}
\usepackage{balance}

\def\BibTeX{{\rm B\kern-.05em{\sc i\kern-.025em b}\kern-.08em
		T\kern-.1667em\lower.7ex\hbox{E}\kern-.125emX}}

\newcommand{\toolname}{\emph{Code Critters}\xspace}
\newcommand{\toolnamereg}{Code Critters\xspace}

\newcommand{\summary}[2]{%
	\vspace{-0.2cm}%
	\begin{center}%
		\colorbox{gray!20}{%
			\parbox{\linewidth}{%
				\textbf{\textsf{Summary (\textit{#1})}:}~%
				#2%
			}%
		}%
	\end{center}%
}

\begin{document}
	
	\title{Engaging Young Learners with Testing Using the Code Critters Mutation Game}
	
	\author{\IEEEauthorblockN{Philipp Straubinger}
		\IEEEauthorblockA{\textit{University of Passau} \\
			Passau, Germany}
		\and
		\IEEEauthorblockN{Lena Bloch}
		\IEEEauthorblockA{\textit{University of Passau} \\
			Passau, Germany}
		\and
		\IEEEauthorblockN{Gordon Fraser}
		\IEEEauthorblockA{\textit{University of Passau} \\
			Passau, Germany}
	}
	
	\maketitle
	
	\begin{abstract}
		Everyone learns to code nowadays. Writing code, however, does not go
		without testing, which unfortunately rarely seems to be taught
		explicitly. Testing is often not deemed important enough or is just
		not perceived as sufficiently exciting. Testing \emph{can} be
		exciting: In this paper, we introduce \toolname, a serious game
		designed to teach testing concepts engagingly. In the
		style of popular tower defense games, players strategically position
		magical ``portals'' that need to distinguish between creatures
		exhibiting the behavior described by correct code from those that
		are ``mutated'', and thus faulty. When placing portals, players are
		implicitly testing: They choose test inputs (i.e., where to place
		portals), as well as test oracles (i.e., what behavior to expect),
		and they observe test executions as the creatures wander across the
		landscape passing the players' portals.  An empirical study
		involving 40~children demonstrates that they actively engage with
		\toolname. Their positive feedback provides evidence that they
		enjoyed playing the game, and some of the children even continued to
		play \toolname at home, outside the educational setting of our
		study.
	\end{abstract}
	
	
	\section{Introduction}
	
Programming is not only a useful and desirable skill in
\mbox{industry~\cite{hired,DBLP:conf/issre/StraubingerF23}}, but it is also
considered a core aspect of computational
thinking~\cite{wing2006computational}. Consequently, programming has
become an essential aspect of education, both at
schools~\cite{kafai2014connected} and in higher education.
Writing code is inseparably linked to testing, which often plays an
insignificant role in education~\cite{seth2014organizational}, despite
growing awareness of its importance in higher education and proposals
to integrate testing into school
curricula~\cite{DBLP:conf/acse/Carrington97, jones2001experiential,
	DBLP:conf/iticse/MarreroS05}.
This neglect is often due to a lack of teaching resources and skills
for testing, along with the perception among learners and programmers
that testing is dull and
tedious~\cite{DBLP:conf/issre/StraubingerF23}. Without a more engaging
and accessible approach to teaching testing, this situation is
unlikely to change.

\begin{figure}[tb]
	\centering
	\includegraphics[width=\linewidth]{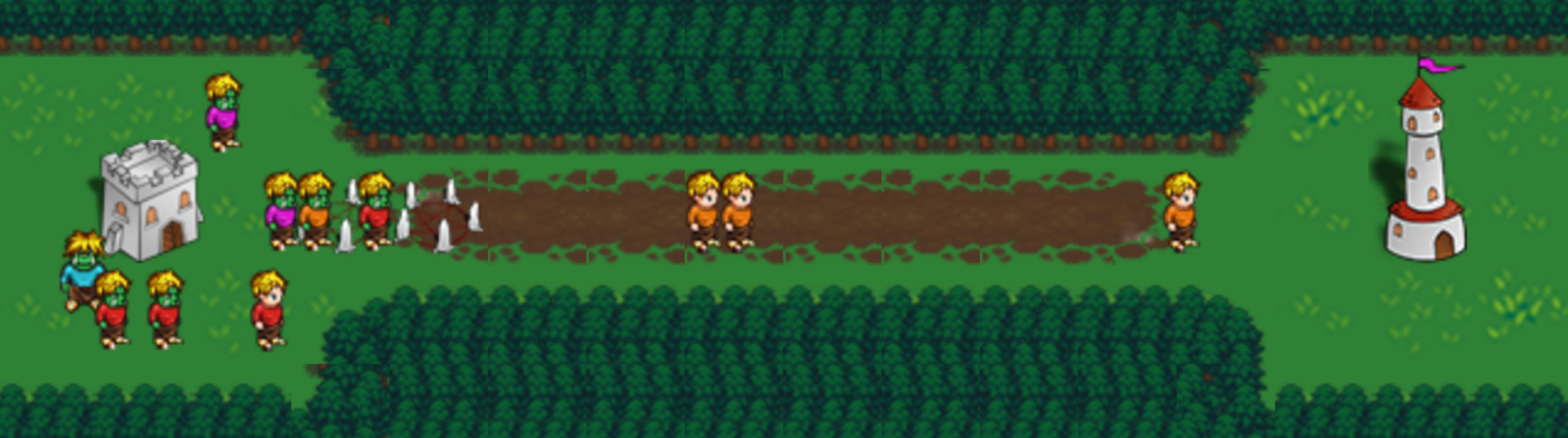}
	\caption{Gameboard during active gameplay}
	\label{fig:play}
\end{figure}

To bridge this educational gap and foster a more engaging approach to
teaching testing, we introduce \toolname, a serious game inspired by
the Tower Defense game genre (see \cref{fig:play}). \toolname is set in a scenario where a
disease threatens humanoid creatures, forcing them to migrate. To help these creatures, the
players are tasked to strategically place magic portals
along the routes taken by them, thereby protecting healthy
creatures from mutated ones, with the ultimate goal of rescuing the
healthy humans and to identify as many mutants while using as few portals as possible.

This competitive process hides an educational aspect: As the creatures
move around the game field, their behavior is determined by small
snippets of simple code, written in an easily understandable
block-based programming language. Mutant creatures are not only
mutated in their appearance but also use mutated, erroneous
code. Thus, distinguishing healthy from mutated creatures resembles
the task of testing programs: By selecting where to place portals,
players implicitly select test data in terms of coordinates or
terrain. Magic portals distinguish healthy critters from mutants using
simple block-based test assertions, thus letting players create test
oracles. By incorporating various software concepts across different
levels and mutant behaviors, \toolname provides an immersive learning
journey for testing concepts.

The game-like nature of \toolname and its use of a simple block-based
representation of code targets younger learners specifically. To evaluate whether \toolname is indeed suitable for younger
audiences, we conducted an empirical study involving 40 children aged 11 to 16. In
particular, we aim to study how children interact with the game
elements, whether their gameplay leads to meaningful testing, and
whether they enjoy the experience. Overall, the contributions
of this paper therefore are as follows:
\begin{itemize}
	\item We introduce the \toolname serious game and its content, the
	idea of which was briefly described in a previous short
	paper~\cite{DBLP:conf/icst/StraubingerCF23}.
	\item We present the results of an empirical study with 40 children in
	a school context, tasked to play the game.
\end{itemize}

Our experiment shows that the majority of children are actively engaged
with \toolname, revealing either a deep interest in understanding each
level thoroughly, or aiming to swiftly progress to the final stage. A
subsequent exit survey confirms that the children enjoyed playing
\toolname, with some even choosing to continue playing during their
free time.

	\section{Background}
	Despite its significance and the increasing recognition in higher education, software testing remains inadequately addressed in programming instruction~\cite{DBLP:journals/jss/GarousiRLA20}. One promising strategy to encourage developers to write tests is gamification, which involves integrating game elements such as leaderboards, points, or challenges into non-game \mbox{contexts~\cite{DBLP:conf/icse/StraubingerF22,DBLP:conf/icse/StraubingerF23,DBLP:conf/mindtrek/DeterdingDKN11}}. Serious games take this a step further, as they are explicitly designed for training, education, or simulation, such that players learn about a topic through embedded information without feeling like they are learning or working~\cite{DBLP:conf/chi/RaybournB05}. Despite this potential, only a few serious games have been proposed for software testing~\cite{DBLP:conf/icer/MiljanovicB17, DBLP:conf/icse/PrasetyaLMTBEKM19, DBLP:conf/fie/ToledoLS22}.

Mutation testing, exemplified by its integration into the Code Defenders~\cite{DBLP:conf/sigcse/FraserGKR19} or Code Immunity Boost games~\cite{hsueh2023design}, stands out as a testing concept well-suited for gamification and serious games. In mutation testing, artificial defects are introduced into the code under test to reveal weaknesses in existing tests~\cite{5487526}. Small variations of the code under test, known as mutants, are created, and the available tests are run against them. If a test fails, it signifies that the mutant has been detected (i.e., \emph{killed}), while mutants remain alive if there are no failing tests,
indicating potential deficiencies in the test suite. In Code Defenders~\cite{DBLP:conf/sigcse/FraserGKR19}, this process is gamified: attackers create artificial defects (mutants), and defenders attempt to detect them by writing tests. However, a significant drawback of Code Defenders and similar gamified testing approaches is that they demand reasonably advanced programming skills, making them more suitable for higher education. To engage less experienced learners, testing needs to be introduced earlier in programming education and in a more accessible way.

A common strategy for making programming accessible to younger learners is using block-based languages like Scratch~\cite{maloney2010scratch}. Instead of typing textual code, learners piece together predefined code blocks by dragging and dropping them, quickly creating games and programs. Given the success of block-based programming~\cite{DBLP:journals/cacm/BauGKST17}, this paper explores the idea of similarly lowering the entry barrier for software testing by employing a block-based programming approach.

	\section{Code Critters}

\toolname is a web application available as open source, and also
accessible openly at \mbox{\url{https://code-critters.org}}, designed
to be played directly in the web browser.

\subsection{Game Scenario}

The inhabitants of the mysterious forest, known as the
\toolname~\cite{DBLP:conf/icst/StraubingerCF23}, have enjoyed peaceful
coexistence in their secluded land for an extended period. However,
their tranquility is disrupted when a sudden outbreak of disease
afflicts the \toolname colony. The infected mutants undergo a
disturbing transformation, deviating from their usual behavior and
posing a threat to the colony. Faced with this crisis, the remaining
healthy critters must seek safety by evacuating their city and making
their way to a secure tower located across the expansive forests.
The game board (see
\cref{fig:play}) illustrates this dire scenario, which follows the classic Tower Defense format. In
this strategic game, players are tasked with defending the tower
against infected critters, ensuring that only the uninfected ones
successfully reach safety.

\subsection{Game Concept}

\begin{figure}[t]
	\centering
	\includegraphics[width=0.4\linewidth]{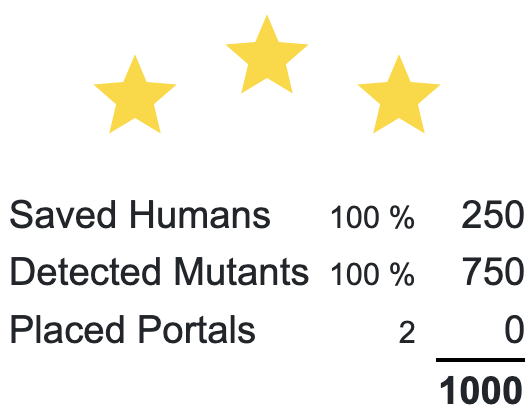}
	\caption{Scoreboard after finishing the current level}
	\label{fig:scorePopup}
\end{figure}

\toolname includes different levels, where each level consists of a
playing field made of 16 by 16 tiles with different types of
terrain and textures: grass, dirt, ice, water, or wood. Critters can
traverse only on grass, dirt, and ice fields. Two tiles on the board
designate the colony’s village (spawn point) and the tower as their
destination, connected by at least one path for the critters to walk
on. The game starts with a playing field consisting only of the
terrain and devoid of portals, with critters stationed in their
village.

As the critters traverse the playing field from village to tower,
healthy critters have a humanoid appearance, whereas some of
the critters are infected and can be easily identified by their
mutated appearance. To succeed in the game, players need to
strategically place magical portals along the possible routes to
prevent mutants from reaching the tower, while ensuring that healthy
critters arrive safely. With a target audience of younger learners,
the notion of enemies and elimination common in tower defense games is
reframed as collecting infected critters with a portal and
transporting them to a safe place until a cure for the disease is
found.

Once the player is done positioning portals, the game can commence,
and critters start their journey from the village, navigating slightly
randomized routes toward the tower while passing through
player-placed portals. The game allows for pausing, speeding up, and
resetting at any time. A correctly configured portal efficiently beams
away mutants while permitting healthy critters to pass without
interference.

A level is completed once all critters have left the village and have
either arrived in the tower or have been collected by portals. The
player then receives a score that is calculated based on how many
healthy critters made it safely and how many mutant critters were
collected, for a maximum of 1000 points, given 20 critters per
level. Each level has a predefined number of portals that is required
to identify all mutants and for each additional portal required by
the player there is a penalty of 25 points.
Finally, the overall score is also rated with 1--3 stars depending on
the point range.
\Cref{fig:scorePopup} illustrates the score popup after 
completing level one. Achieving the maximum score of 1000 for this
level involves placing two portals, one on grass and one on the dirt
trail.

\subsection{Critter Behavior}

\begin{figure}
	\centering
	\includegraphics[width=\linewidth]{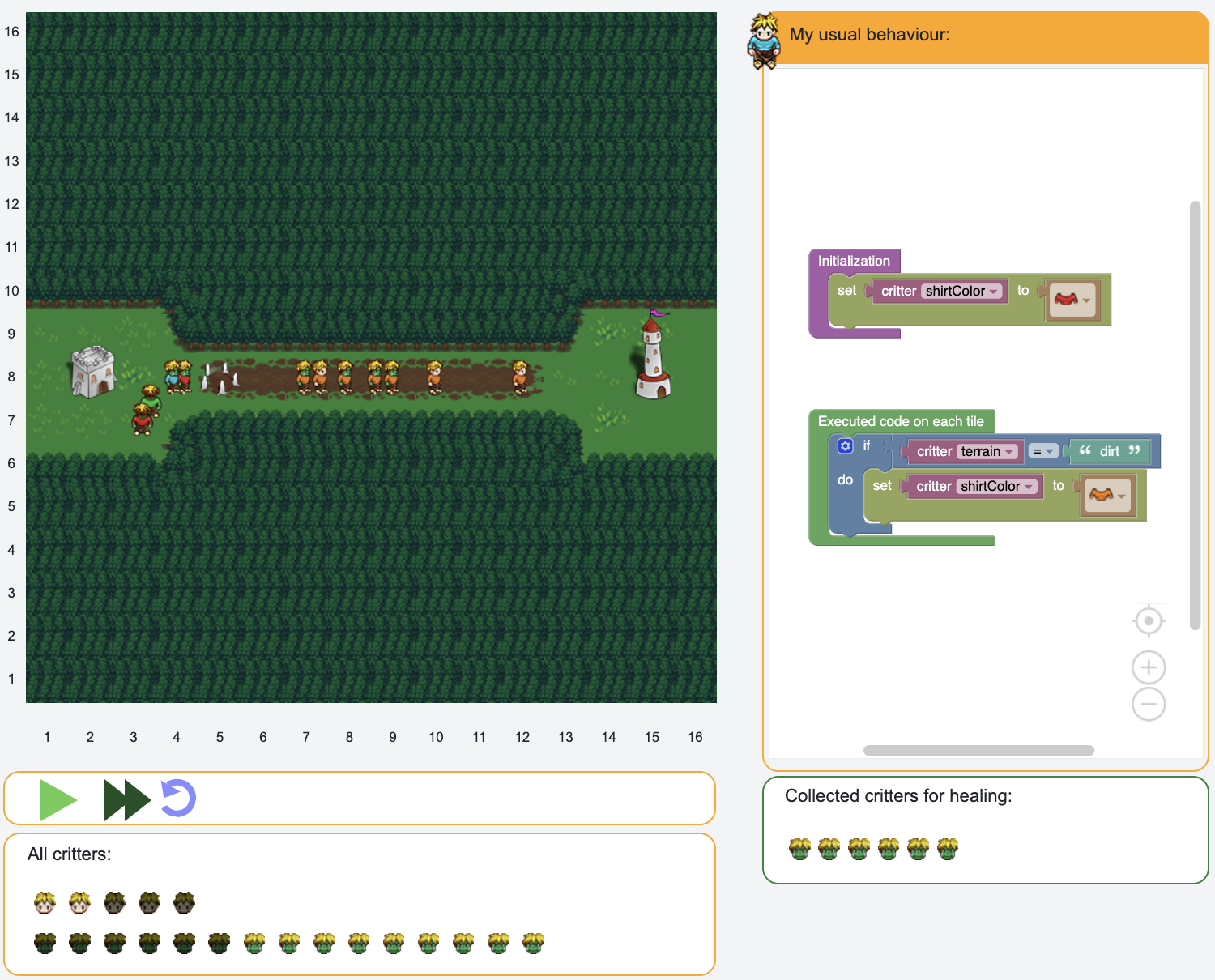}
	\caption{Game screen of \toolname}
	\label{fig:game}
\end{figure}

The game interface, as depicted in \cref{fig:game}, features the main
gameboard on the left and game control buttons at the bottom. The
initially empty green box on the right gradually fills up with
collected critters throughout the game. To the right of the gameboard
%
the appearance and behavior of this level's critters is described as the Critter
Under Test (CUT), which serves as a concise set of instructions in the
form of code that is invoked continuously as critters traverse the
landscape. To make the game more accessible, \toolname uses the
Blockly\footnote{https://developers.google.com/blockly} library to
visually represent code as blocks rather than text.

\begin{figure}
	\centering
	\includegraphics[width=\linewidth]{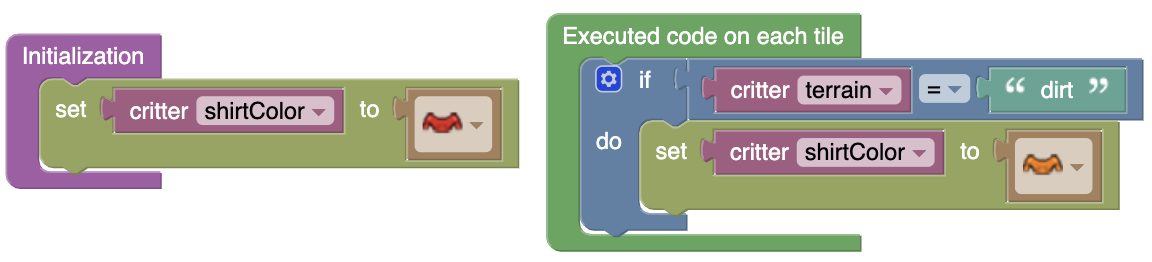}
	\caption{Critter under test}
	\label{fig:healthyCut}
\end{figure}

Conceptually, a CUT can be likened to functions within an
object-oriented class, outlining both the attributes and behavior of
a critter.
\Cref{fig:healthyCut} shows a closer view of the CUT corresponding to
the game in \cref{fig:game}. The \textit{Initialization} section
establishes initial values for attributes such as shirt or hair
color. The \textit{Executed code on each tile} represents a looping
mechanism executed repeatedly on each tile. For instance, the CUT in
\cref{fig:healthyCut} initializes the critter with a red shirt, and
the color changes when the critter steps on a dirt field.

\begin{figure}
	\centering
	\includegraphics[width=\linewidth]{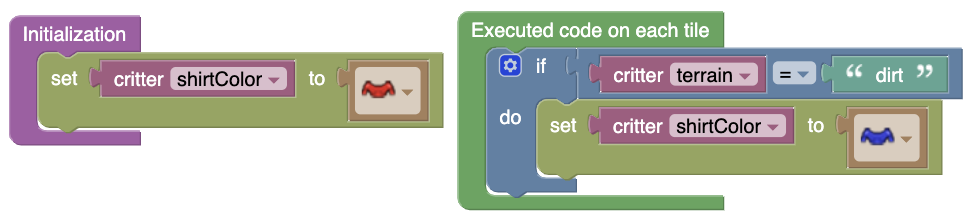}
	\caption{A mutant of the CUT in \cref{fig:healthyCut}}
	\label{fig:mutantCut}
\end{figure}

Mutant critters execute code that deviates from the CUT's code,
similar to the artificial defects used in mutation testing. As a point
of comparison, \cref{fig:mutantCut} illustrates the CUT of an infected
critter: A mutant deviates in one or more attributes from the correct
CUT, indicating the presence of at least one code mutation. Various
versions of mutants exist within the game. In the example of
\cref{fig:mutantCut}, the mutant begins with a red shirt (correct) but
subsequently changes to a blue shirt upon reaching a dirt field.

\subsection{Portals as Critter Tests}

\begin{figure}[t]
	\centering
	\includegraphics[width=\linewidth]{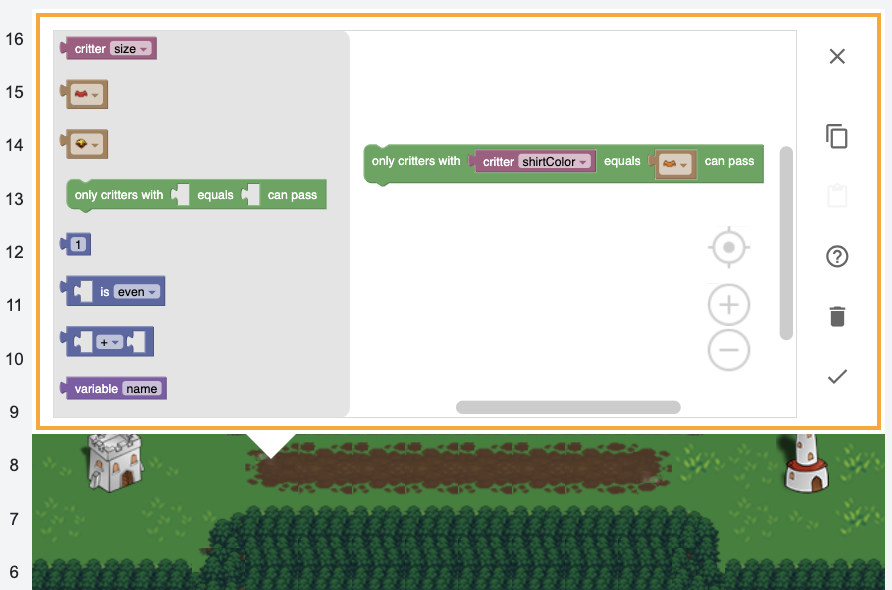}
	\caption{Gameboard of \toolname with an open portal}
	\label{fig:gameWithPortal}
\end{figure}

The overarching objective is to ensure that only critters exhibiting
correct behavior reach the tower while capturing mutants along the
way. This can be accomplished by strategically placing portals along
the critters' path. \Cref{fig:game} shows an example
portal positioned at the start of the dirt trail, resembling the
depiction in \cref{fig:gameWithPortal} when opened. Each portal
occupies a single tile and functions as a test, with the tile
properties (coordinates and texture) serving as test inputs. The test
specified in the portal is executed during gameplay whenever a critter
steps onto it.

To set up a test case, players can click on any walkable field on the
board, prompting a dialog to appear with a toolbox of programming
blocks. Constructing a test case involves initiating the process with
the \textit{only critters with ... can pass} block, akin to a familiar
assert statement. Through judicious combination with other blocks and
asserting the appropriate attributes, the portal becomes instrumental
in safeguarding the tower. The cumulative effect of all placed portals
constitutes the comprehensive test suite for the CUT, serving as the
tools the player employs to thwart the enemies.

As an example, consider the mutant in \cref{fig:healthyCut}, transitioning from a red
to a blue shirt on dirt. Placing a portal on the first dirt tile
aligns with the if-condition of the correct CUT, where the shirt color
mutation occurs. As healthy critters turn orange at this tile, the
portal's code, exemplified in \cref{fig:gameWithPortal}, should assert
the expected shirt color as orange. This ensures that only
orange-dressed critters pass, successfully collecting blue-shirted
mutants. The visualization consists of the portal emitting a
beaming light, and the captured mutant soaring across the
screen into the designated collection space.

\subsection{Systematic Testing with Portals}

The portal in \cref{fig:gameWithPortal} on its own is not sufficient
to thwart \emph{all} mutants from reaching the tower in
\cref{fig:game}. Since the disease induces varying mutations, testing
all behaviors and attributes becomes crucial. Considering
\cref{fig:healthyCut}, another statement remains uncovered by the
portal in \cref{fig:gameWithPortal}: the Initialization. Placing an
additional portal at the beginning of the grass path, asserting the
expected red shirt color, completes the test suite. Although there is
no limit on the number of portals, minimizing the test suite is
encouraged, with unnecessary and redundant portals resulting in
deducted points from the score.

\begin{figure}
	\centering
	\includegraphics[width=\linewidth]{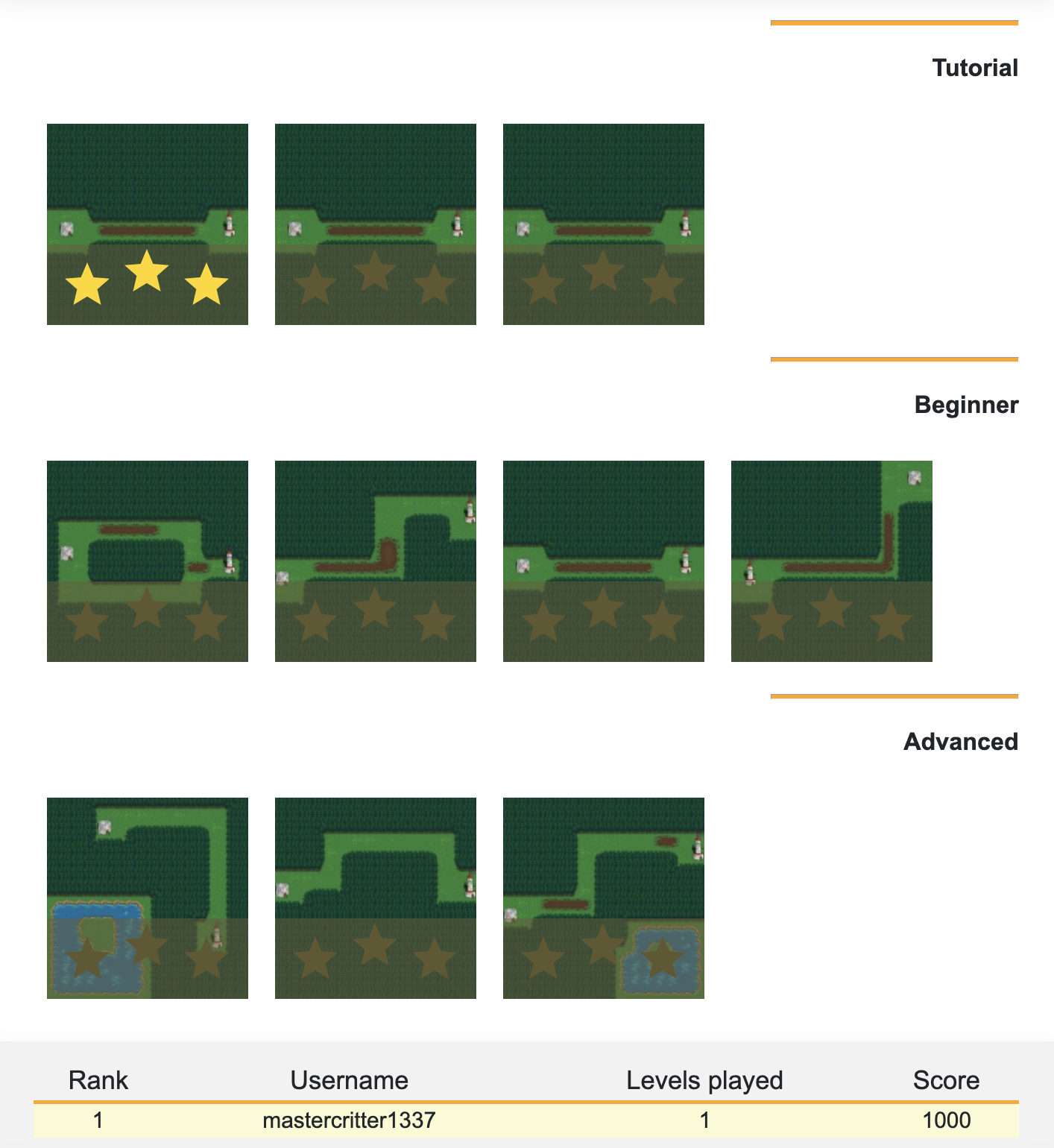}
	\caption{Different levels in \toolname with the leaderboard}
	\label{fig:levels}
\end{figure}

Since players need to systematically test the code of CUT, each level
can be used to exercise different testing concepts, such as statement
or branch coverage. Currently, \toolname incorporates a total of ten
levels, categorized into three groups: three tutorial, four beginner,
and three advanced levels (refer to \cref{fig:levels}). The
progression in difficulty is marked by the introduction of more
intricate CUTs and the introduction of new concepts, including the use
of named variables. The complexity is further enhanced by altering the
design of the gameboard itself, such as incorporating additional and
varied terrain fields or creating multiple possible routes to the
tower.  Cumulative scores from all played levels are tallied and
ranked in the score leaderboard, visible at the bottom in
\cref{fig:levels}.


	\section{Evaluation}
	To evaluate the usefulness of \toolname, we conducted a controlled experiment, aimed at answering the following research questions:

\begin{itemize}
	\item RQ 1: How do children play \toolname?
	\item RQ 2: How much testing do players of \toolname do?
	\item RQ 3: Do children enjoy playing \toolname?
\end{itemize}

\subsection{Experiment setup}

The controlled experiment was carried out at the Maristengymnasium Fürstenzell in two sessions in January 2024.

\subsubsection{Experiment Environment}

Thanks to a collaboration with Maristengymnasium Fürstenzell, a secondary school that prepares students for higher education at universities, we conducted our experiment within their academic setting. Specifically, we were granted access to two slots within their elective subject centered on robotics. Each slot entailed 90 minutes of teaching, with approximately 20 students participating in each session. Given that this course is an elective subject, all participating students had expressed an interest in programming robots. Before the experiment, these students had three months of experience programming robots using a block-based language developed by Lego,\footnote{\url{https://education.lego.com/en-gb/lessons/ev3-robot-trainer/}} making them already familiar with the foundational concept of block-based programming, which forms the foundation of \toolname.

None of the students had engaged in any explicit form of testing before the experiment, relying primarily on trial and error~\cite{edwards2004using}. In total, 40 students took part in our experiment across the two slots, with the majority being male and only a single female participant. Most participants were in the 5th to 7th grade (11--13 years old), with only four students in the 8th to 10th grade (14--16 years) classes.

\subsubsection{Experiment Procedure}

In the initial ten minutes, we provided an introduction to \toolname, elucidating its storyline and game mechanics. Subsequently, participants were assigned the task of independently playing \toolname, without specific instructions on which level to tackle or whether to accumulate all points before progressing to the next level. This gameplay phase extended for 60 minutes, concluding with the cessation of the experiment, allowing ample time for participants to respond to our exit survey.

\subsubsection{Experiment Analysis}

Our analysis primarily focuses on presenting results for all participants and groups of them. To assess the significance of measurements between different groups we use the exact Wilcoxon-Mann-Whitney test \cite{10.1214/aoms/1177730491} to calculate $p$-values with $\alpha = 0.05$. When visualizing trends, we display the 84.6\% confidence intervals for mean calculations~\cite{fisher1956statistical}. If the intervals overlap, indicating no statistically significant difference~\cite{DBLP:journals/csda/AfshartousP10}, it is equivalent to confirming non-significance using an exact Wilcoxon-Mann-Whitney test with $\alpha = 0.05$.

\subsubsection{RQ 1: How do children play \toolnamereg}

To address this research question, we examine the data collected throughout the experiment. Firstly, we analyze the average number of (1) completed and (2) attempted levels, along with the corresponding achieved (3) scores and (4) stars. Secondly, we investigate the participants' activity during the experiment, searching for differences in their behavior. This includes identifying the levels they played at different times during the experiment and the number of games played within specific time intervals during the experiment.

\subsubsection{RQ 2: How much testing do players of \toolnamereg do}

To address this research question, we conduct a comparative analysis of the total number of (1) generated test cases (i.e., portals), (2) identified bugs (i.e., mutants), and (3)~recognized correct code (i.e., healthy critters) across the participants. Additionally, we examine these metrics during specific time intervals, precisely every minute of the experiment, to gain insights into the development of participants' testing skills over time. Our comparison does not involve absolute values; rather, we focus on the ratios between (1)~utilized and required portals, (2) killed and total mutants, and (3) finished and total healthy critters. Furthermore, we conduct these ratio comparisons on a per-level basis rather than per player to uncover differences in the difficulty levels. To ascertain statistical significance, we employ the exact Wilcoxon-Mann-Whitney test with $\alpha = 0.05$ to compare between groups of participants.

\subsubsection{RQ 3: Do children enjoy playing \toolnamereg}

This question is answered by comparing the answers to the exit survey. The survey encompasses general inquiries about gender and grade, followed by a section with seven five-point Likert scale questions asking participants' enjoyment of different aspects of \toolname, along with open-ended prompts inviting them to share additional thoughts or feedback. For better visualization, the data is presented in stacked bar charts including the questions and percentages.

\subsection{Threats to Validity}

\paragraph{Threats to Internal Validity} Participants' prior experience with block-based programming could potentially bias the experiment's outcomes. However, introducing \toolname without prior exposure to block-based programming might overwhelm the children. Additionally, participants may feel inclined to provide socially desirable responses in the exit survey, potentially skewing the data. To counter this threat, we instructed them to respond honestly and without hesitation.

\paragraph{Threats to External Validity} The small sample size and lack of diversity in terms of gender and grade levels limit the generalizability of our findings to broader populations of children. Moreover, since the children were part of a robotics course and already had an interest in programming, the results may not fully represent the experiences of children in a more general context. Furthermore, the short duration of the experiment may not capture the long-term effects or usage patterns of \toolname, potentially impacting how interactions with the tool develop over time.

	\section{Results}
	\subsection{RQ 1: How do children play \toolnamereg?}

\begin{figure*}[t]
	\centering
	\begin{subfigure}[t]{0.325\textwidth}
		\centering
		\includegraphics[width=\textwidth]{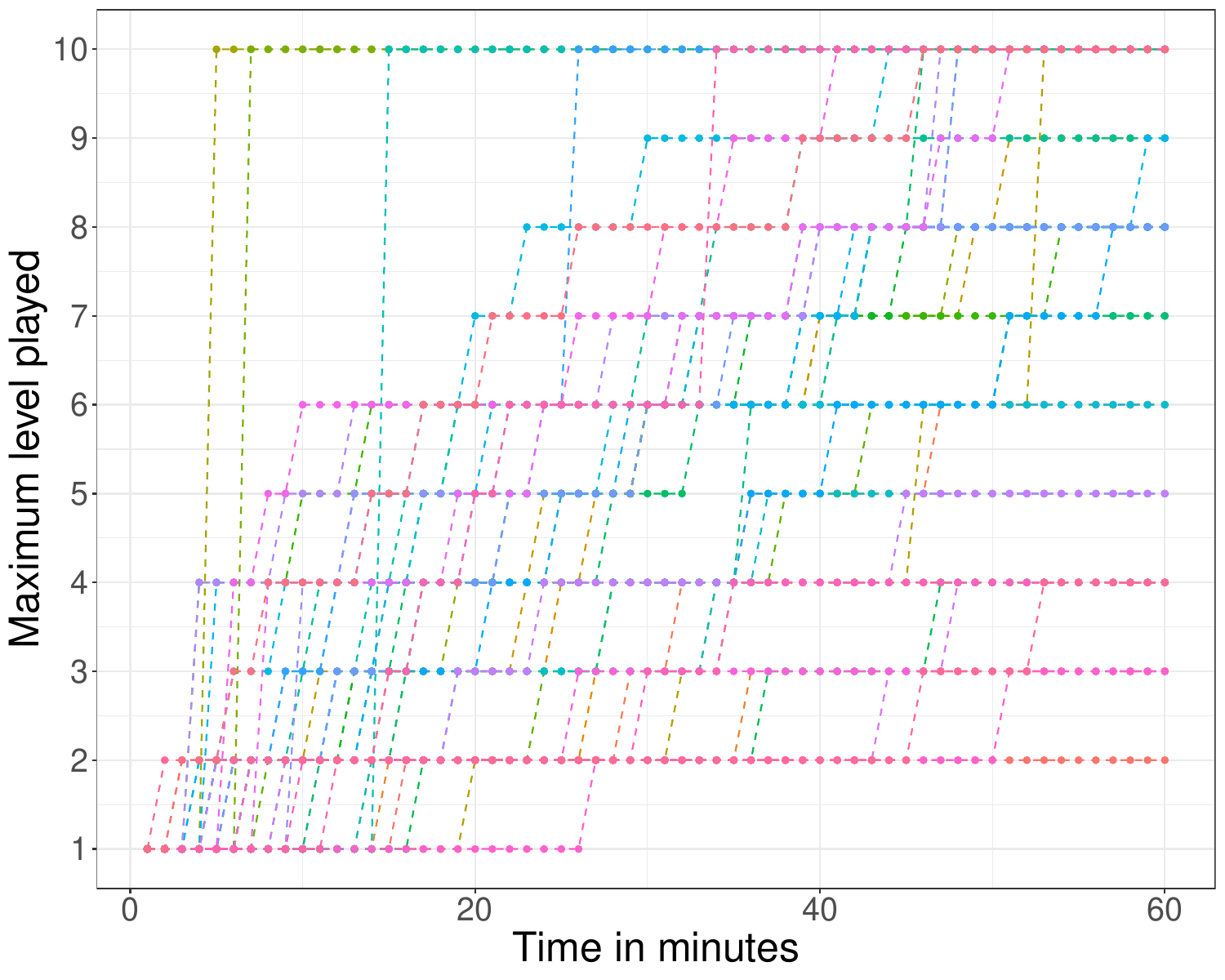}
		\caption{Maximum level played over time}
		\label{fig:timemaxlevel}
	\end{subfigure}
	\hfill
	\begin{subfigure}[t]{0.325\textwidth}
		\centering
		\includegraphics[width=\textwidth]{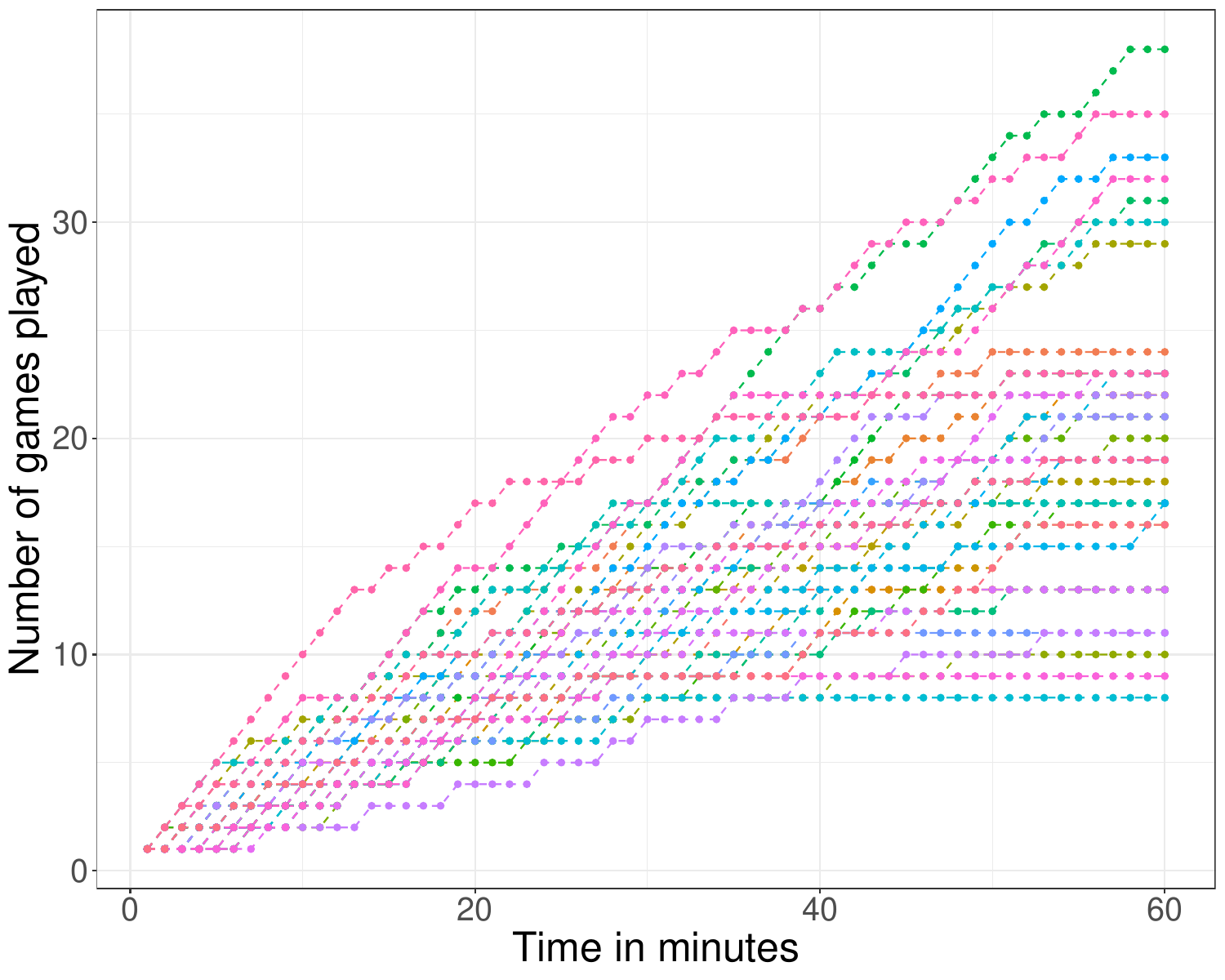}
		\caption{Number of games played over time}
		\label{fig:timegames}
	\end{subfigure}
	\hfill
	\begin{subfigure}[t]{0.325\textwidth}
		\centering
		\includegraphics[width=\textwidth]{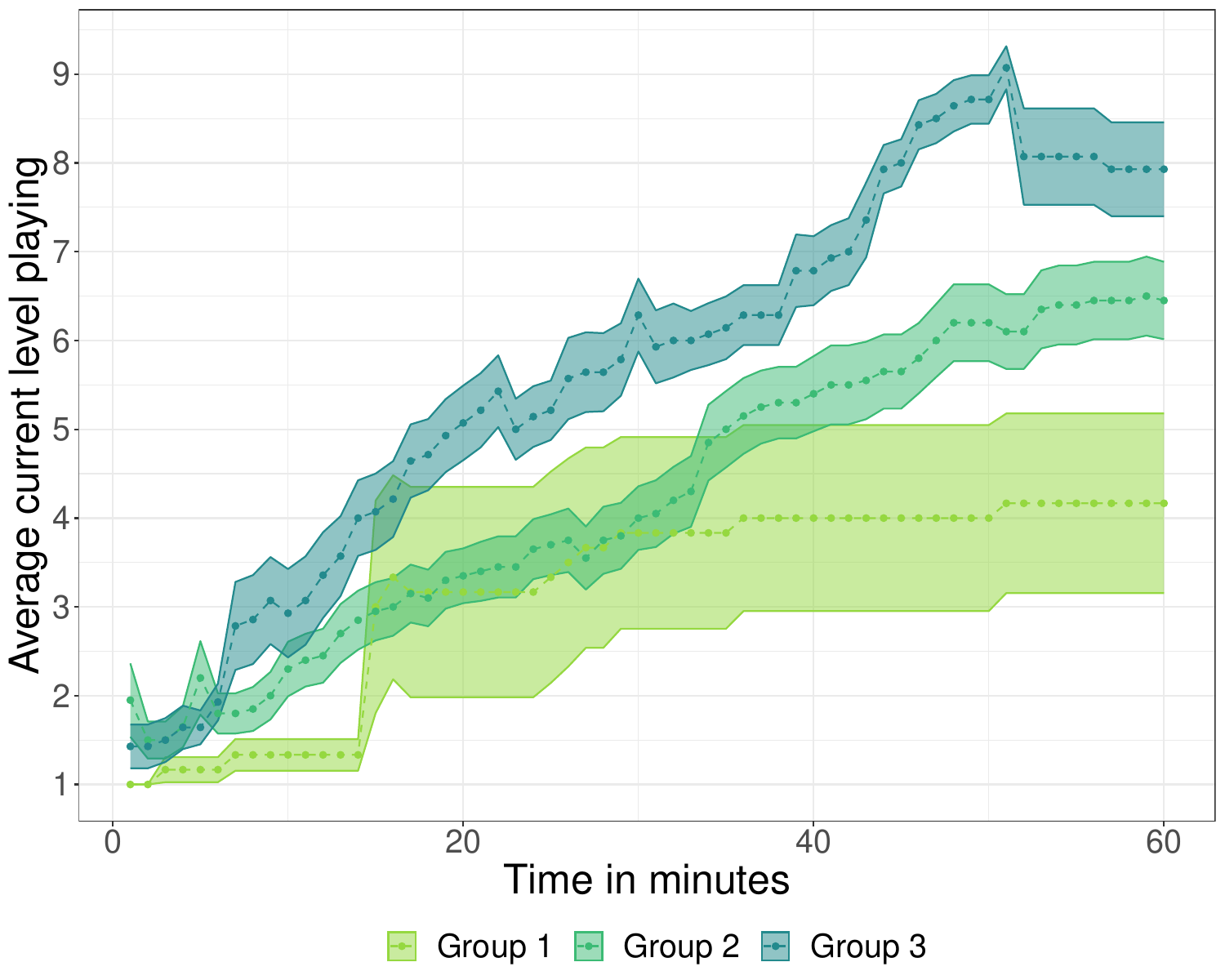}
		\caption{Current level played over time divided into groups}
		\label{fig:timelevel}
	\end{subfigure}
	
	\caption{Differences between the players over time}
	\label{fig:difftime}
\end{figure*}

During the experiment, the children altogether played 824 games, with an average of 20.6 games per child. They explored around 6.42 distinct levels, finishing approximately 6.12 levels each with at least one star. Achieving at least one star constituted completing a level, prompting some participants to proceed to the next level, aiming to finish it. On average, they scored 5153 points, accumulating a total of 12.68 stars. It took them about 7 minutes, on average, to complete a level. During the levels, the players placed a mean of about 50 portals in total, while identifying more than 200 mutants and saving more than 40 healthy critters. 


Considering performance over time, especially achieved levels, reveals substantial discrepancies and variations between players. \Cref{fig:timemaxlevel} shows the progression in terms of levels over time for all participants and reveals a wide range of maximum achieved levels from two to ten. The number of games played over time (\cref{fig:timegames}) similarly shows discernible differences among the children, suggesting the presence of three distinct groups. This observation aligns with our observations during the experiment, where we noticed three distinct groups of children exhibiting different behaviors while playing \toolname:

\begin{itemize}
	\item Group 1: These children demonstrated minimal participation in the experiment, often placing random portals without deliberate strategy, and observing the behavior of mutants and healthy critters.
	\item Group 2: These children displayed a meticulous approach, striving to comprehend the intricacies of \toolname and consistently aiming for 100\% completion of each level before progressing.
	\item Group 3: These children prioritized speed, seeking to complete all levels, advancing to the next level as soon as they achieved at least one star in the previous one.
\end{itemize}

Each of these approaches leads to a different number of levels completed, and we can therefore categorize the children into groups by the number of different levels completed: 1--3 for Group 1 (6 children), 4--7 for Group 2 (20 children), and 8--10 for Group 3 (14 children). Analyzing the average current level played during the experiment (\cref{fig:timelevel}), we observe significant differences between these groups, with non-overlapping intervals in the latter third of the experiment. Group 1 shows minimal progress throughout the experiment, indicating limited achievement. In contrast, Group 2 demonstrates consistent progress, steadily completing one level after another. Group 3 exhibits rapid advancement, reaching their maximum level by minute 51. Subsequently, many of them progress to level 10 before revisiting lower levels to improve their scores.

\summary{RQ 1}{We observed three main types of behavior: A few
	children disengaged from the game, most children tried to
	progress as quickly as possible, and many others also focused on
	maximizing their scores.}

\subsection{RQ 2: How much testing do players of \toolnamereg do?}

\begin{figure*}
	\centering
	\begin{subfigure}[t]{0.325\textwidth}
		\centering
		\includegraphics[width=\textwidth]{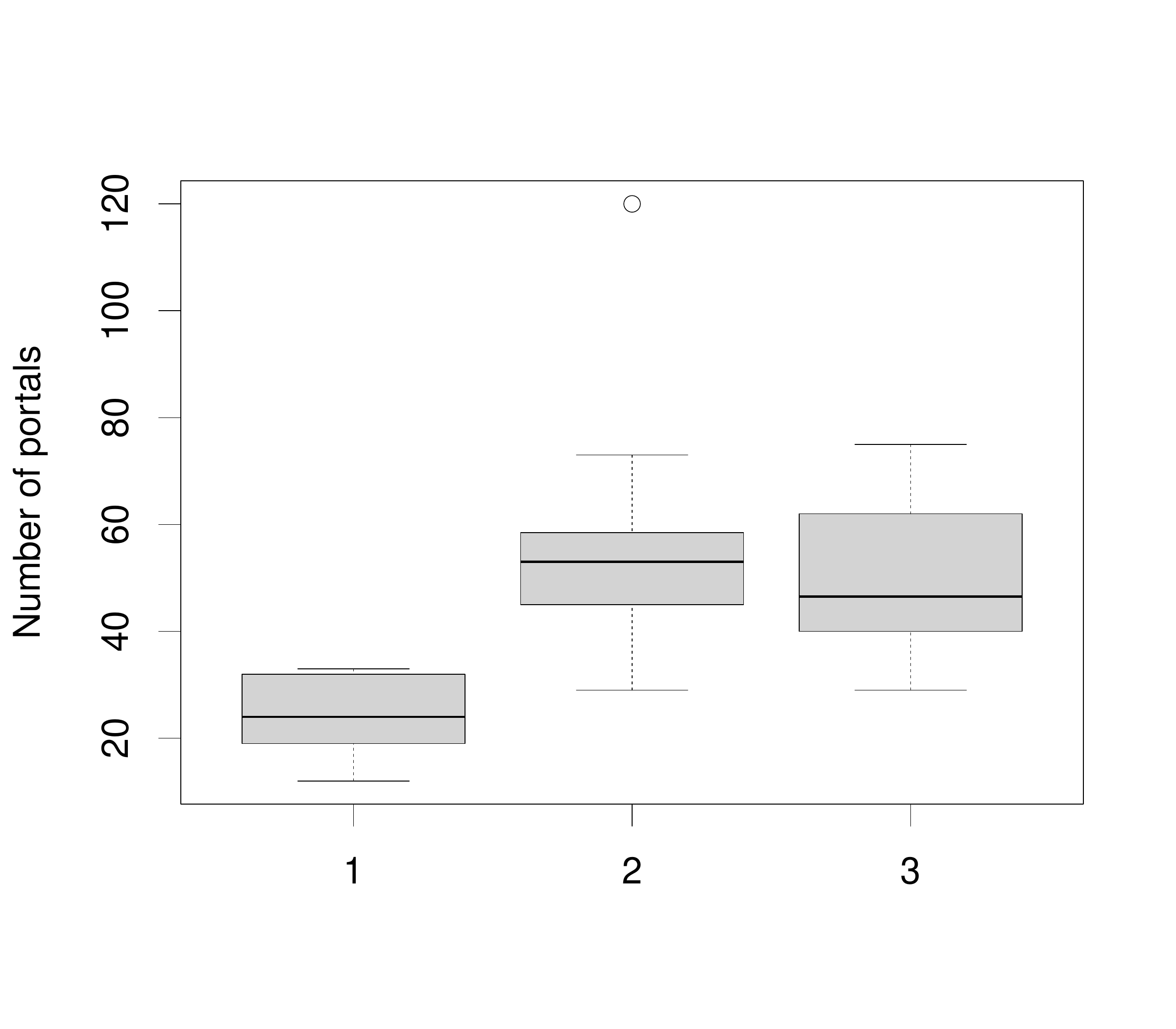}
		\vspace{-3.5em}
		\caption{Number of created portals}
		\label{fig:boxmines}
	\end{subfigure}
	\hfill
	\begin{subfigure}[t]{0.325\textwidth}
		\centering
		\includegraphics[width=\textwidth]{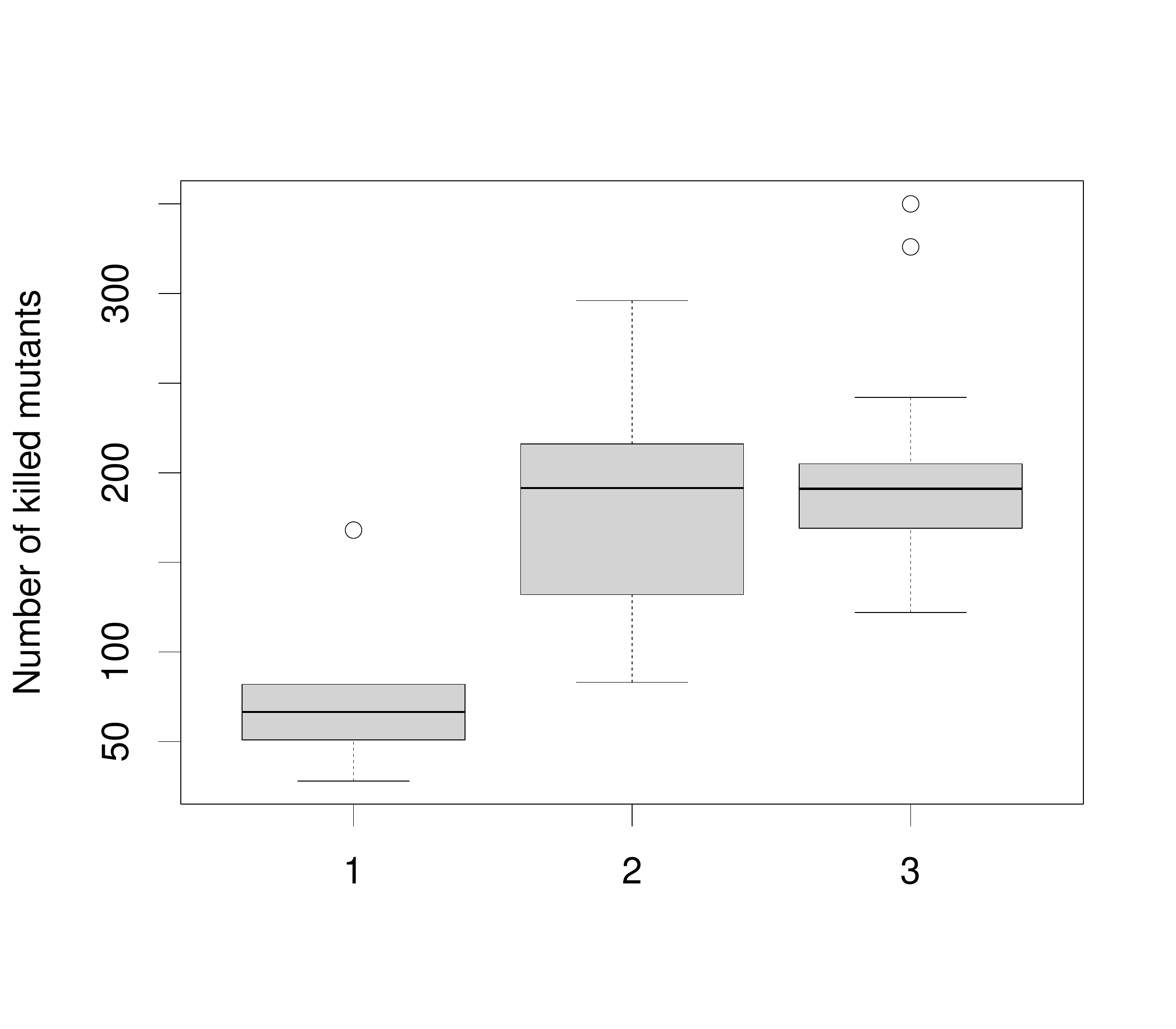}
		\vspace{-3.5em}
		\caption{Number of killed mutants}
		\label{fig:boxmutants}
	\end{subfigure}
	\hfill
	\begin{subfigure}[t]{0.325\textwidth}
		\centering
		\includegraphics[width=\textwidth]{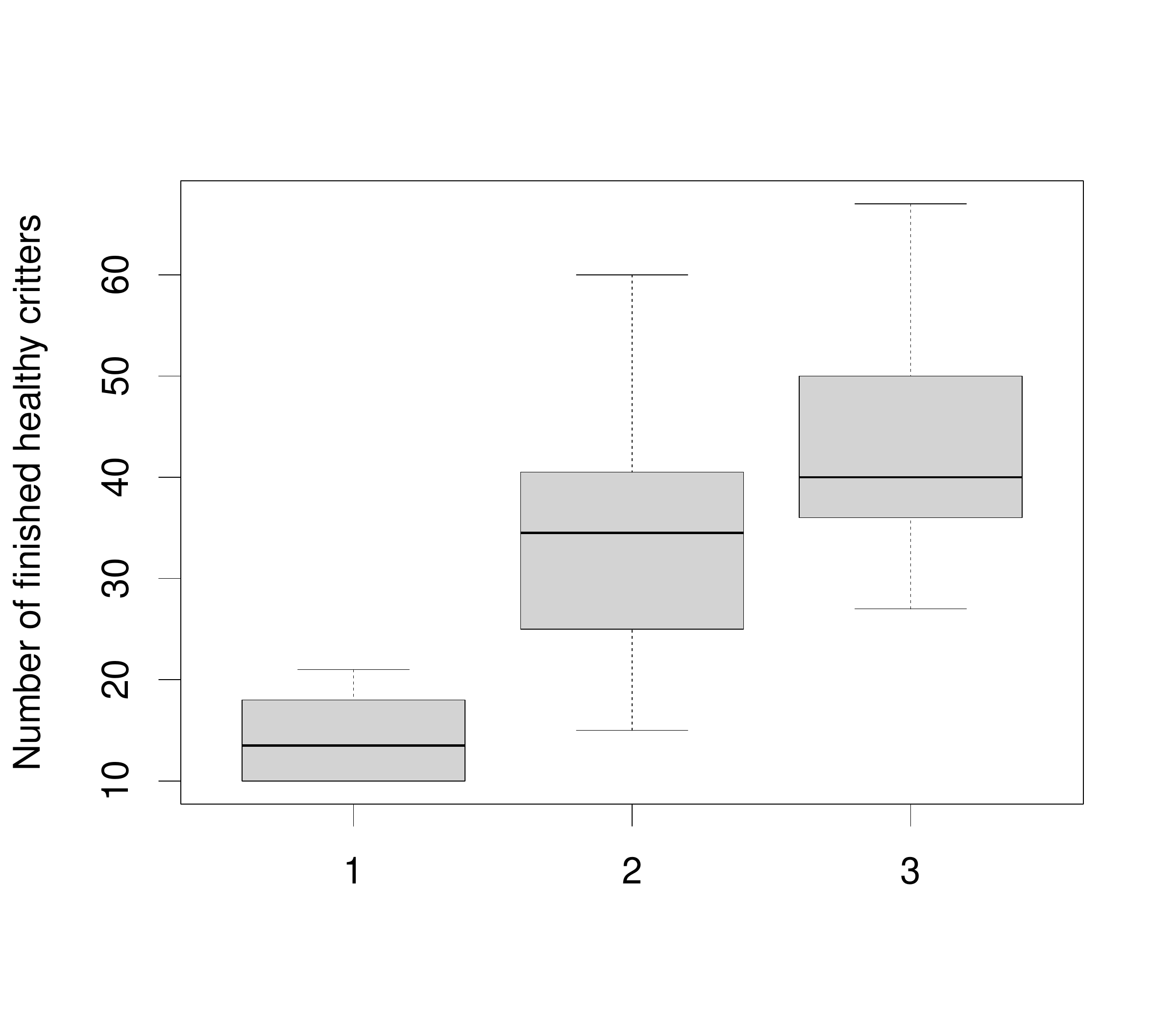}
		\vspace{-3.5em}
		\caption{Number of finished healthy critters}
		\label{fig:boxhumans}
	\end{subfigure}
	
	\caption{Statistics on the use of \toolname divided into groups}
	\label{fig:testboxes}
\end{figure*}

\Cref{fig:boxmines} shows the number of portals created by each group
identified in RQ1.  Participants of Group 1 seem substantially less
ambitious (avg. 24.00): Both Groups 2 (54.35) and 3 (50.57)
significantly outperform Group 1 in the number of created portals
($p<0.001$ for both). However, there is no significant difference
between Groups 2 and 3.

To understand how effective the portals the players created are, we
can consider how many mutants were detected, shown per group in
\cref{fig:boxmutants}. The result resembles that of portal creation,
with a notable difference between Group 1 and both Groups 2
($p=0.002$) and 3 ($p<0.001$). This underscores the lower
participation level of Group 1, while the outcomes of the other two
groups are comparable.

A complementary data point is whether the portals successfully let
healthy critters pass, or falsely identified them as
mutants. \Cref{fig:boxhumans} again shows that both Group 2 ($p=0.002$)
and 3 ($p<0.001$) significantly outperform Group 1. However, in this
comparison, there is a more pronounced difference between Group 2 and
3, with a $p$-value of $0.061$, close to $\alpha = 0.05$. This can be
explained by children in Group 3 aiming to maximize their scores at
individual levels, thus focusing more on portal placement to recognize
correct behavior.

\begin{figure*}
	\centering
	\begin{subfigure}[t]{0.325\textwidth}
		\centering
		\includegraphics[width=\textwidth]{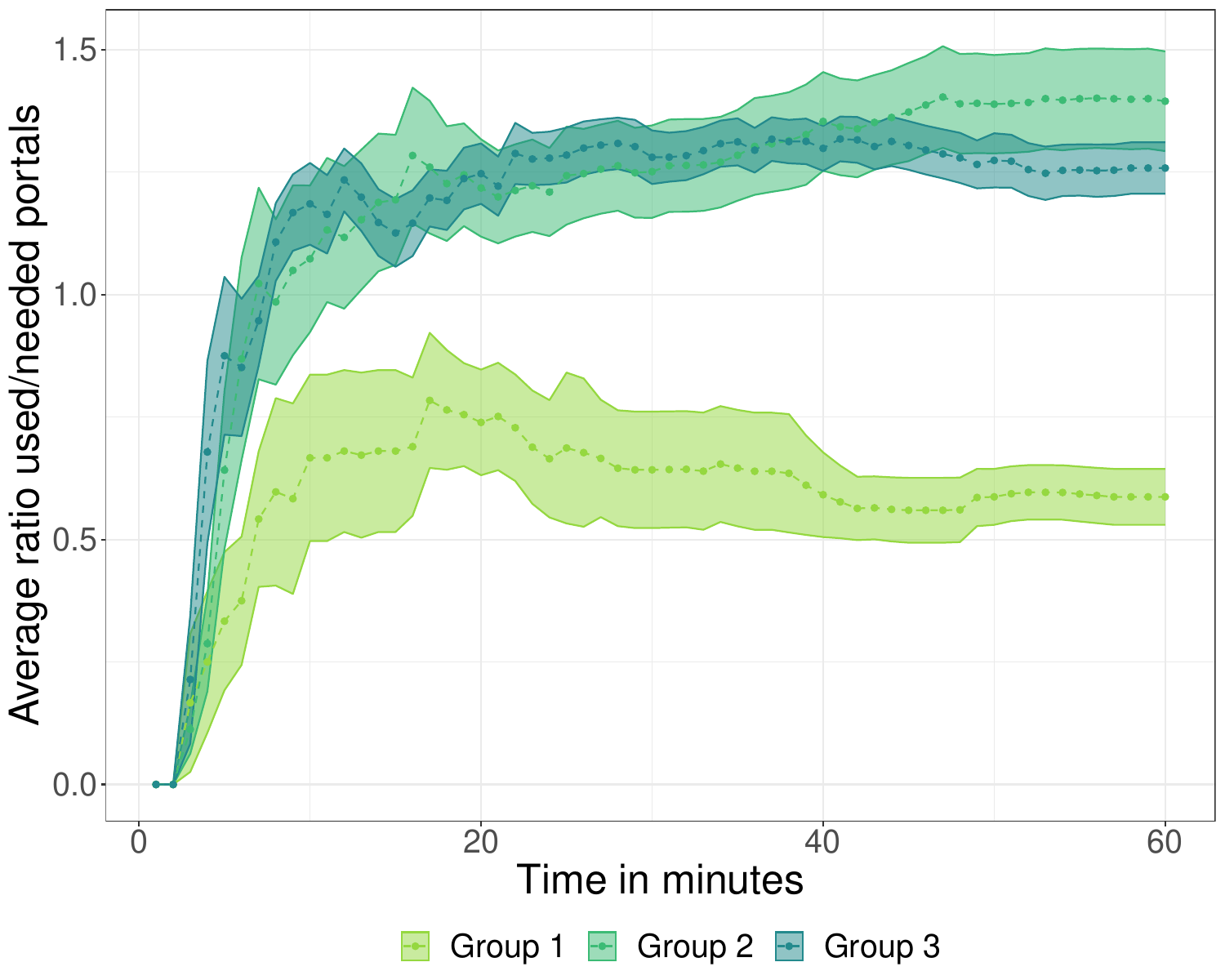}
		\caption{Average ratio between used and needed portals over time}
		\label{fig:timeminesneeded}
	\end{subfigure}
	\hfill
	\begin{subfigure}[t]{0.325\textwidth}
		\centering
		\includegraphics[width=\textwidth]{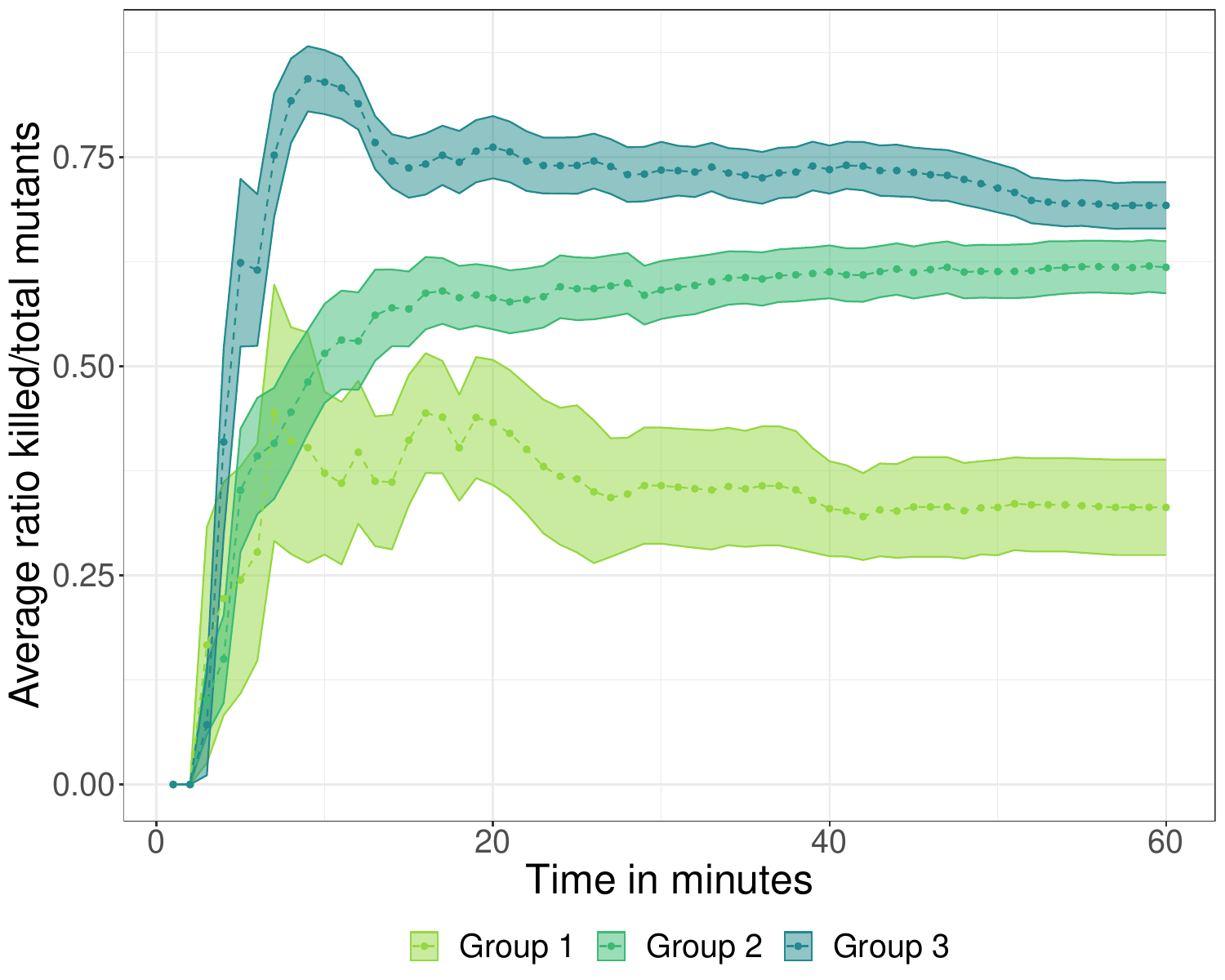}
		\caption{Average ratio between killed and total mutants over time}
		\label{fig:timemutants}
	\end{subfigure}
	\hfill
	\begin{subfigure}[t]{0.325\textwidth}
		\centering
		\includegraphics[width=\textwidth]{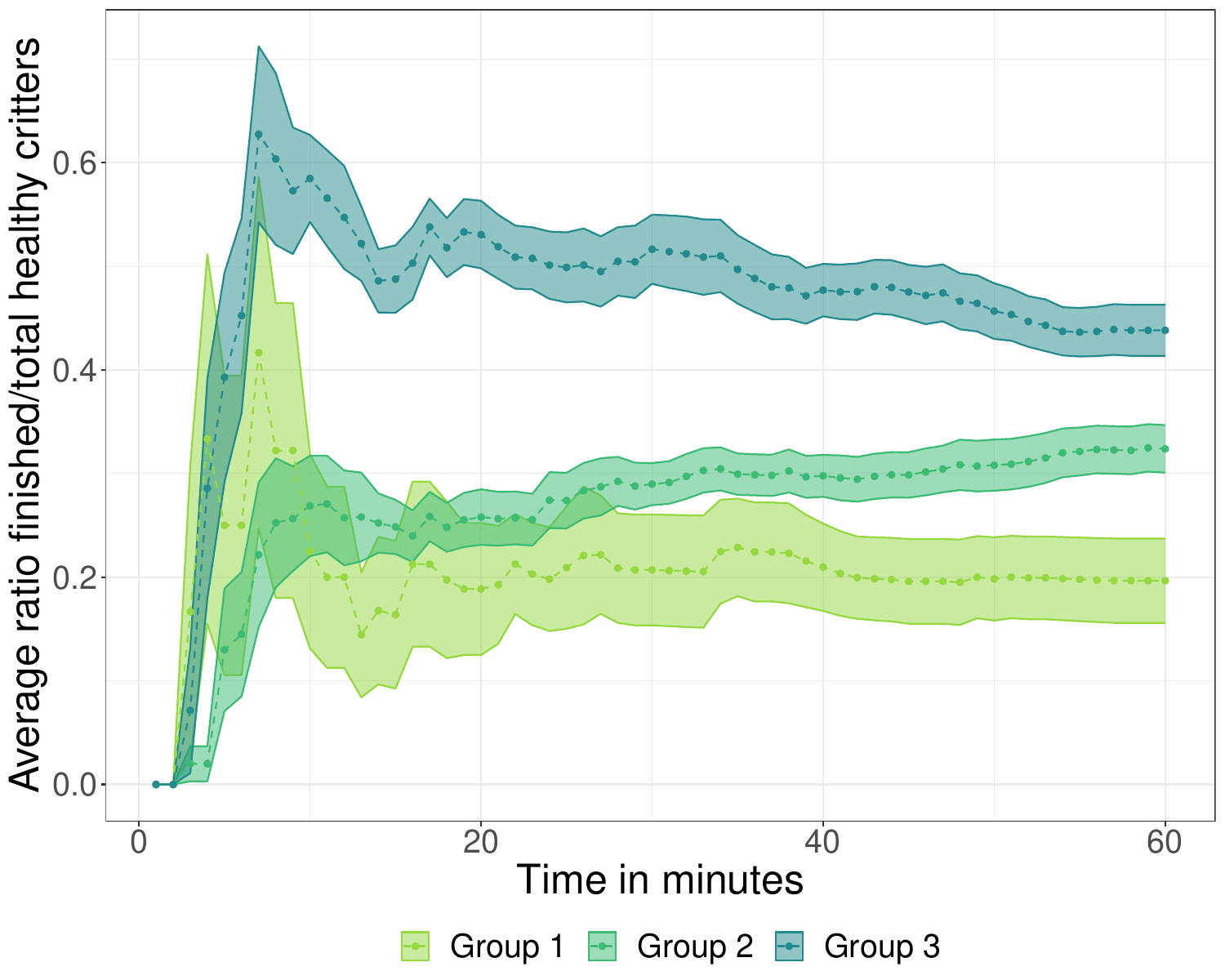}
		\caption{Average ratio between finished and total healthy critters over time}
		\label{fig:timehumans}
	\end{subfigure}
	
	\caption{Differences in correctness between the groups over time}
	\label{fig:diffcorrecttime}
\end{figure*}

\begin{figure*}
	\centering
	\begin{subfigure}[t]{0.49\textwidth}
		\centering
		\includegraphics[width=\textwidth]{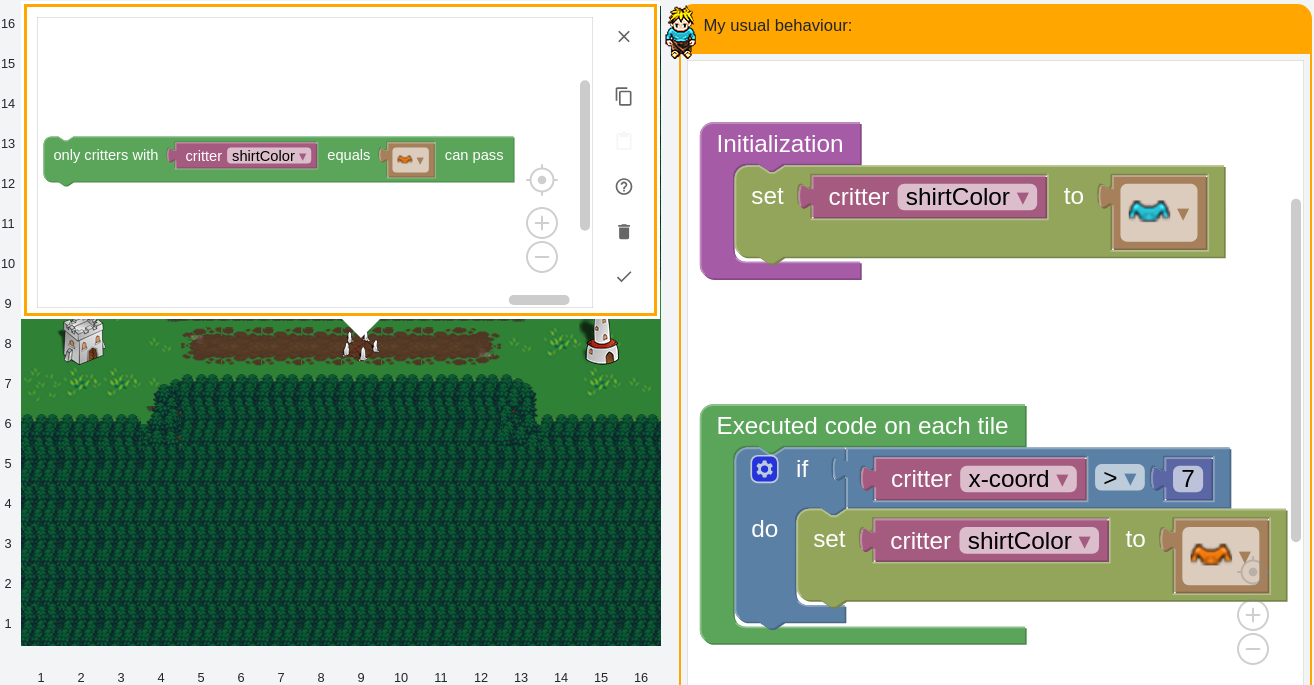}
		\caption{Misplaced portal in level 2}
		\label{fig:level2}
	\end{subfigure}
	\hfill
	\begin{subfigure}[t]{0.48\textwidth}
		\centering
		\includegraphics[width=\textwidth]{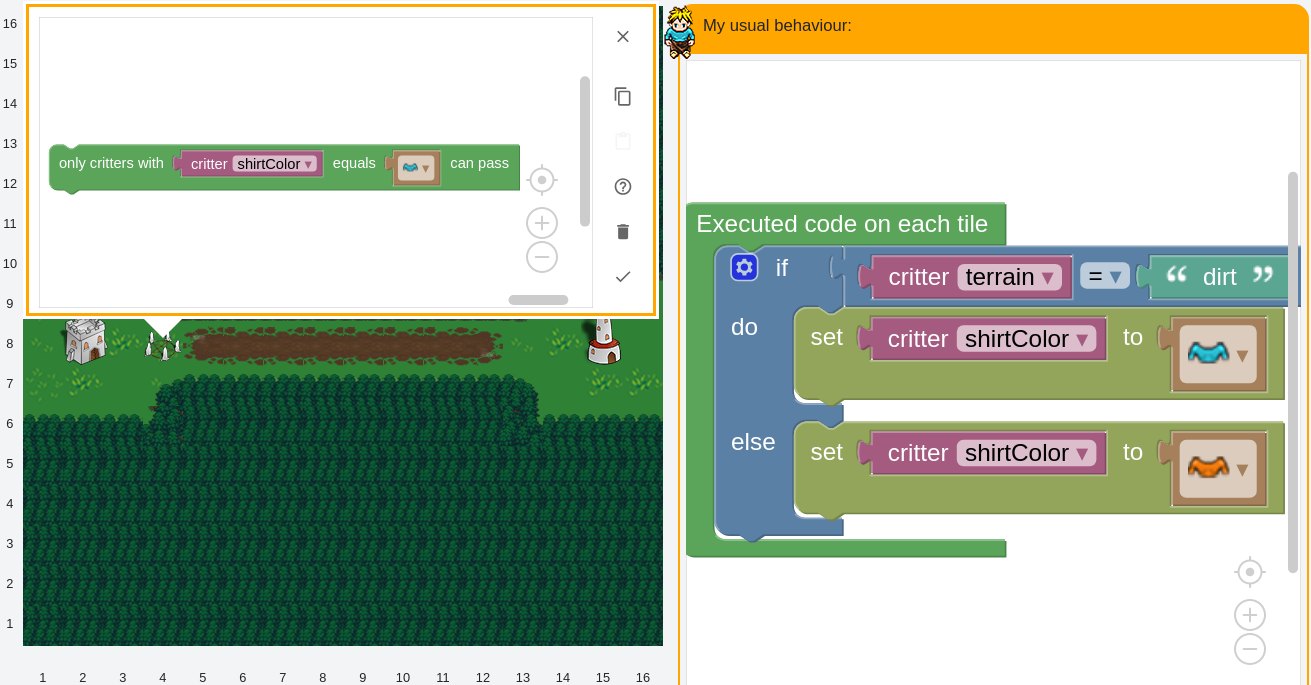}
		\caption{Misplaced portal in level 3}
		\label{fig:level3}
	\end{subfigure}
	
	\caption{Two examples of misplaced mines}
	\label{fig:misplacedlevels}
\end{figure*}

\Cref{fig:timeminesneeded} compares the number of portals used with
the number of portals required to solve a level as an indicator of
effectiveness. Participants of Group 1 used insufficiently many
portals, suggesting that they struggled to identify all the relevant
cases they needed to check for in the CUT's behavior. The behavior of
Groups 2 and 3 is generally similar in that they used more portals
than actually necessary to win a level, resulting in a ratio greater
than 1, but thus also higher coverage of the CUT's behaviors. However,
towards the end, Group 3's ratio noticeably begins to decline towards
the desired ratio of 1, suggesting an improved performance. We
conjecture that this is influenced by the score penalty for redundant
portals.

Clear differences in behavior can also be identified by considering
the ratio of eliminated mutants to the total number of mutants
(\cref{fig:timemutants}) and the ratio of completed healthy critters
to their total number (\cref{fig:timehumans}).
The average ratio of detected mutants does not surpass 90\%, which can
be attributed to various factors, primarily stemming from misplaced or
erroneous portals. For instance, consider \cref{fig:level2}, which
illustrates such a scenario. As per the CUT, the healthy critter
changes its shirt color to orange at x coordinates of 8 and
above. Placing the portal on the tile with x coordinate 9 renders the
code susceptible to mutants incorrectly altering their shirt color at
an x coordinate of 8, consequently evading detection.

Similarly, the average ratio of surviving healthy critters does not
surpass 70\%, such that more than 30\% were erroneously eliminated on
a portal. This could stem from various factors, primarily arising from
misinterpretation of conditions or misunderstanding of how assertions
function. For instance, consider \cref{fig:level3}, which depicts such
a scenario. Here, an if-condition with two branches is presented. The
children mistakenly believed that the shirt color changed from orange
to blue on the dirt instead of the opposite. Consequently, they placed 
portals allowing only blue shirts to pass on grass instead of dirt,
resulting in incorrect elimination of healthy critters.

Overall, while Group 3 excels in terms of mutants identified,
\cref{fig:timemutants} as well as \cref{fig:timehumans} both show a
small but steady decline over time. This is likely due to these
players progress to more challenging levels quicker
(\cref{fig:timelevel}). In contrast, players in Group 2 focused more
on perfecting each level, progressing slowly, and repeatedly placing
similar portals until correctly positioned, often resulting in healthy
critters casualties, but also exhibiting a steady increase in both
ratios.
Group 1 members were initially motivated but eventually resorted to
randomly placing portals, which may explain the initially good and
later deteriorating ratios. The differences between groups are
significant for the ratio of eliminated mutants
(\cref{fig:timemutants}) Towards the end, particularly the disparities
between Group 1 and both Groups 2 ($p=0.001$) and 3 ($p<0.001$) are
significant, while the difference between Groups 2 and 3 diminishes,
nearing significance ($p=0.051$).
Around two-thirds into the experiment, significant differences emerge
between all groups in terms of the ratio of completed healthy critters
(\cref{fig:timehumans}), with $p<0.031$ between Group 1 and 2, and
($p=0.024$) between Groups 2 and 3.

\begin{figure}[t]
	\centering
	\includegraphics[width=0.75\linewidth]{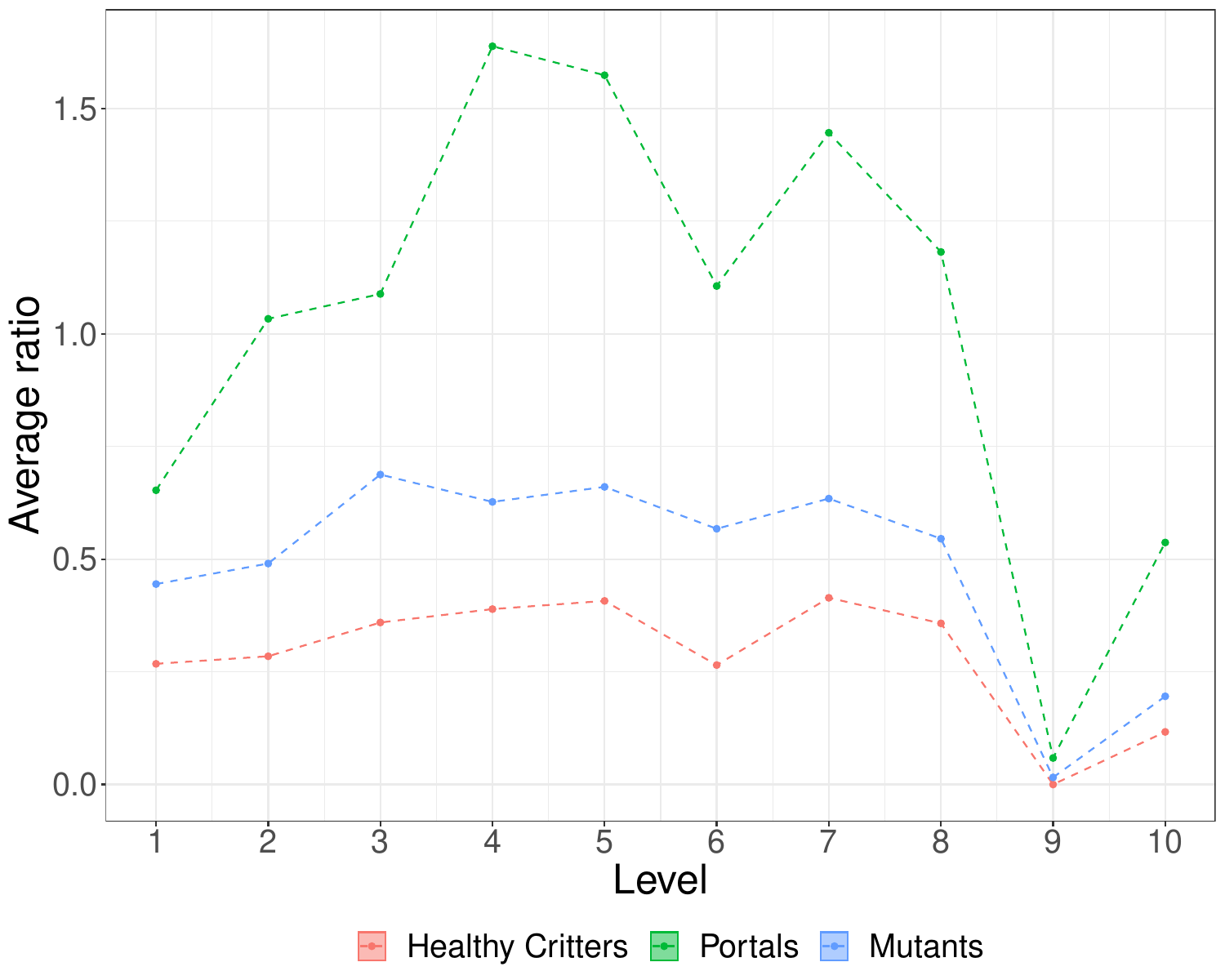}
	\caption{Correctness per level}
	\label{fig:correctnesslevel}
\end{figure}

\begin{figure}[t]
	\centering
	\includegraphics[width=0.9\linewidth]{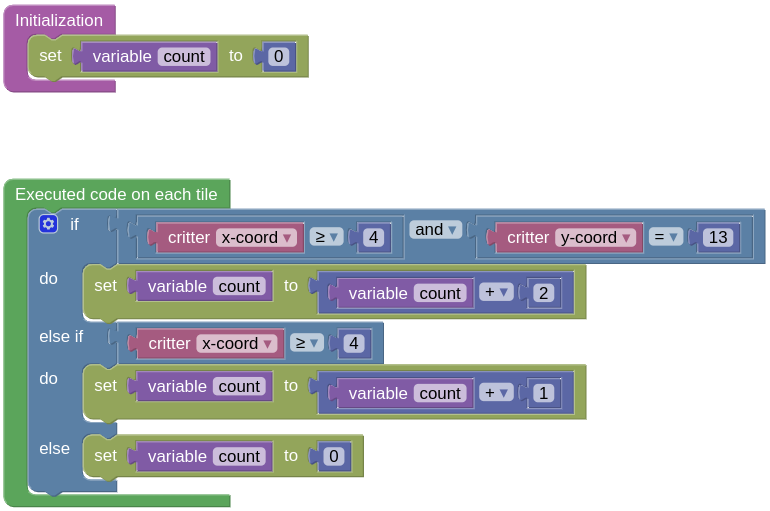}
	\caption{CUT of level 9}
	\label{fig:level9}
\end{figure}

The influence of individual levels can also be seen in
\cref{fig:correctnesslevel}, which compares the ratios of portals,
mutants, and healthy critters per level.  The ratio of detected
mutants and surviving healthy critters remain relatively consistent
across levels 1 through 8, while there is an increase in the portal
ratio for levels 4, 5, and 7. In the case of level 4, one possible
factor of influence may be that there are multiple possible paths to
the tower; in addition levels 4 and 5 use a disjunction of two
conditions in an if-statement, which likely also contributes to
uncertainty concerning necessary tests. Level 7 is the first one
to include an if-if/else construct, which also seems to cause some
uncertainty about necessary tests with players.
Level 6 adds complexity by introducing variables, resulting in a
slight decrease in the mutation score and surviving healthy critters,
but since it only uses a simple if-condition without an else-branch and
the level offers only one path, it seems to be more obvious to players
how many portals are needed.

A clear outlier is represented by level 9 (\cref{fig:level9}), where
all ratios drop to nearly 0.0, identifying it as the most challenging
level. In this level, the terrain does not influence the behavior of
the critters, but there is a nested if-if-else construct that changes
a variable. The code does not affect the behavior or properties of the
critters at all, and assertions would be necessary purely on the value
of the variable. Although the code in level 10 is similarly complex,
there is no variable and this time the critters change their
appearance, thus allowing assertions on critter attributes rather than
variables; this seems to be slightly easier to comprehend for
children.

\summary{RQ 2}{The children engaged well with testing actions in our
	experiment, creating effective tests (portals) that detected an
	average of 90\% of mutants and 70\% of healthy
	critters. Complexities in level design as well as code lead to
	variation in accuracy as well as redundant tests.}

\subsection{RQ 3: Do children enjoy playing \toolnamereg?}

\begin{figure}
	\centering
	\includegraphics[width=0.9\linewidth]{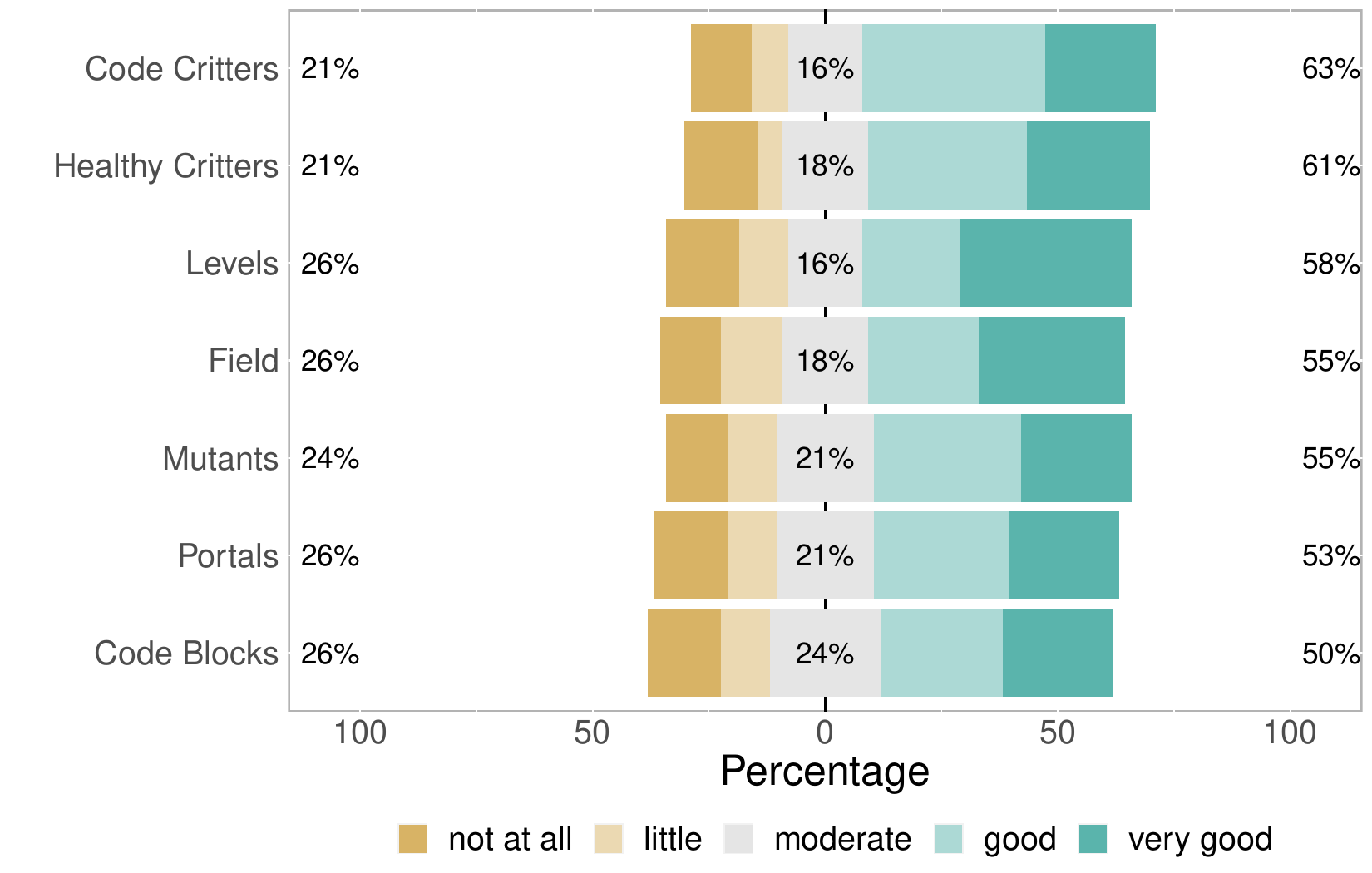}
	\caption{Survey responses to ``How much did you enjoy\ldots''}
	\label{fig:survey}
\end{figure}

\Cref{fig:survey} summarizes the results of the exit
survey. According to the responses, 63\% of
the children found \toolname enjoyable. They particularly enjoyed the
healthy critters and levels, while feeling more neutral towards portals and
code blocks. Interestingly, no specific aspects of the game were
disliked by a significant portion of the children, as at least half
of them expressed enjoyment for every question and therefore every part of the game. Notably,
17.5\% of the children continued playing \toolname at home in their
free time, indicating its overall appeal as an enjoyable game.

\summary{RQ 3}{The children enjoyed playing \toolname, even continuing to play it in their free time.}

	\section{Conclusions}
	\toolname is a serious game designed to teach children software
testing concepts through block-based mutation testing in an enjoyable
manner. Our study involving 40 children demonstrates their active
engagement with \toolname, as they successfully identified numerous
bugs and often correctly identified the behavior of the correct
code. Feedback from the exit survey indicates that the children
thoroughly enjoyed playing \toolname, with many continuing to play
even after the experiment concluded.

Moving forward, we aim to expand the game by incorporating additional
programming concepts such as loops to enhance their understanding of
testing. Furthermore, we plan to implement a hint system to assist
children struggling to progress in the game.  In addition, our goal is
to bridge the gap between block-based programming languages and more
sophisticated ones such as Python.  We are considering to achieve this
by introducing additional levels in which block-based code is
gradually replaced by another programming language while maintaining
consistent gameplay.

\vspace{1em}
\noindent The source code of \toolname is available at:

\begin{center}
	\href{https://github.com/se2p/code-critters}{https://github.com/se2p/code-critters}
\end{center}

\noindent You can try out \toolname online at:

\begin{center}
	\href{https://code-critters.org}{https://code-critters.org}
\end{center}


	\section*{Acknowledgements}
	\noindent This work is supported by the DFG under grant \mbox{FR 2955/2-1}, ``QuestWare: Gamifying the Quest for Software Tests''. Thanks to Laura Caspari for her contributions to developing \toolname, and Philipp Unger for supporting the empirical study.
	
	\balance
	\bibliographystyle{IEEEtran}
	\bibliography{bib}
	
\end{document}